\documentclass[a4paper,twosides,twocolumn]{article}
\usepackage{multicol,multirow,graphics,epstopdf}
\usepackage{amsmath,amsfonts,amssymb,bm,upgreek,mathrsfs,mathcomp}
\usepackage{multicol}
\usepackage{flafter}
\usepackage[pagewise,switch,columnwise]{lineno}
\usepackage[dvipsnames]{xcolor}
\usepackage{graphicx}
\usepackage{caption}
\usepackage{stfloats}
\usepackage{csquotes}
\usepackage{float}
\usepackage{titlesec}
\usepackage{2022}
\usepackage{fancyhdr}

\fancypagestyle{firstpage}{%
  \fancyhf{}%
  \fancyfoot[L]{%
  \parbox{\textwidth}{\footnotesize
    \raggedright
    $^{*}$Corresponding authors. Email: zhufeng25@mail.sysu.edu.cn; luole5@mail.sysu.edu.cn\\[2mm]
    \centering \thepage
  }%
}%
}

\begin{document}
\twocolumn[
\vspace* {0mm} \begin{center}
\large\bf{\boldmath{Symmetry-Enforced Non-Hermitian Jarzynski Equality
in an SU(2)-Rotated Family of Hybrid $\mathcal{PT}$--$\mathcal{APT}$ Systems
}}
\par\vspace{5mm}
\normalsize \rm{}Zongru Yang$^{1}$, Teng Liu$^{1}$, Xiaodong Tan$^{1}$, Feng Zhu$^{1,2,4*}$, and Le Luo$^{1,2,3,4,5*}$
\par\vspace{3mm}\small\sl $^{1}$School of Physics and Astronomy, Sun Yat-sen University, Zhuhai 519082, China 

$^{2}$Shenzhen Research Institute of Sun Yat-Sen University, Nanshan Shenzhen 518087, China

$^{3}$Quantum Science Center of Guangdong-HongKong-Macao Greater Bay Area, Shenzhen 518048, China

$^{4}$Guangdong Provincial Key Laboratory of Quantum Metrology and Sensing, Zhuhai 519082, China

$^{5}$State Key Laboratory of Optoelectronic Materials and Technologies, Sun Yat-sen University, Guangzhou, China 510275
\end{center}
\vskip 1.5mm

\small{ The Jarzynski equality is a cornerstone of nonequilibrium thermodynamics, linking work statistics to equilibrium free-energy differences. Although it has been extensively verified in classical and quantum Hermitian settings, its status in non-Hermitian dynamics remains under debate. Here we show that, in a postselected no-quantum-jump framework, a conditional non-Hermitian Jarzynski equality holds when the transition probabilities obey a parity-exchange symmetry. We study a constructed family of two-level hybrid Hamiltonians formed as linear combinations of parity-time ($\mathcal{PT}$) and anti-parity-time ($\mathcal{APT}$) symmetric terms, and demonstrate using complementary geometric and algebraic arguments that the parity-exchange symmetry persists throughout the corresponding $\mathrm{SU}(2)$-rotated orbit. Relative to previous $\mathcal{PT}$-focused conditional Jarzynski equality results, the advance here is an extension of the symmetry criterion from the isolated $\mathcal{PT}$ endpoint to a broader $\mathcal{PT}$--$\mathcal{APT}$ hybrid family. Experimentally, we implement three representative points, $\theta_k = 0, \pi/4, \pi/2$, in a single trapped $^{171}\mathrm{Yb}^+$ ion and measure the resulting work distributions under cyclic protocols with $\Delta F = 0$, confirming the predicted symmetry criterion at those points. Our results establish a symmetry-based extension of the conditional non-Hermitian Jarzynski relation within this restricted two-level setting.

\par}\vskip 6mm
\noindent\textbf{Keywords:} Jarzynski equality, non-Hermitian dynamics, $\mathcal{PT}$ and $\mathcal{APT}$ symmetry, parity-exchange symmetry, trapped ion 

\vspace{6mm}
]
\thispagestyle{firstpage}
\section{Introduction}
\par In thermodynamics, the first and second laws constrain energy and entropy changes, leading, for example, to Jensen's inequality $\langle W\rangle\geqslant\Delta F$, where $\langle W\rangle$ is the ensemble average of work and $\Delta F$ is the free energy difference \cite{Alicki1979,Adkins1983,Fermi2012}. However, these inequalities do not by themselves provide a complete description of fluctuations in driven, far-from-equilibrium processes. To explain phenomena in these non-equilibrium processes, researchers have turned to the framework of statistical mechanics \cite{Esposito2009}, with two of the most prominent results being the Jarzynski equality \cite{Jarzynski1997,Jarzynski2004} and the Crooks fluctuation theorem \cite{Crooks1998,Crooks1999}. Here we focus on the form of the Jarzynski equality:
\begin{equation}
    \langle e^{-\beta W}\rangle=e^{-\beta\Delta F}=\frac{Z(T)}{Z(0)},
    \label{eq:JE}
\end{equation}
where angular brackets $\langle\cdot\rangle$ denote an average over the ensemble of trajectories, $\beta$ denotes the inverse temperature $(k_B=1)$, and $Z(T)$ and $Z(0)$ are the partition functions at time $T$ and time $0$, respectively. The Jarzynski equality has been extensively validated in both classical \cite{Liphardt2002,Douarche2005,Harris2007}
and quantum systems \cite{Huber2008,An2015,Xiong2018,Smith2018} and can be viewed as an extension of
Jensen's inequality.

Notably in quantum systems, the classical notion of a \enquote{trajectory} is no longer applicable, rendering the evaluation of work a non-trivial task. To address this challenge, the two-point measurement (TPM) method has emerged as a widely adopted and effective approach for quantifying work in quantum systems \cite{Tasaki2000,Mukamel2003,Talkner2007}. Within this framework, the work distribution is defined as \cite{Talkner2007}:
\begin{equation}
    P(W)=\sum_{if}\delta\!\bigl(W - [E_f(T) - E_i(0)]\bigr)P_{fi}P_i.
     \label{eq:dis}
\end{equation}
Here, $i$ and $f$ label the initial and final energy eigenstates, respectively. $E_i(0)$ and $E_f(T)$ denote the corresponding energy eigenvalues of the Hamiltonians at the first and second energy measurements. The Dirac delta function $\delta\!\bigl(W - [E_f(T) - E_i(0)]\bigr)$ selects only those events for which the work satisfies $W = E_f(T) - E_i(0)$. $P_{fi}$ and $P_i$ denote the transition probability from the initial state $i$ to the final state $f$ and the occupation probability of the initial state $i$, respectively. The TPM scheme offers a well-defined and experimentally accessible framework in which the characterization of work is reduced to measuring the probability distribution of energy changes in the quantum system.
\par In recent years, non-Hermitian systems have attracted considerable attention due to their unique and rich physical properties, such as exceptional point dynamics \cite{Ashida2020, Zhang2019}, as well as exotic topological phases and the non-Hermitian skin effect \cite{Yao2018, Lin2023}. Notably, recent trapped-ion experiments have demonstrated rapid progress in non-Hermitian physics, including the tomography, programmable simulation, and experimental witnessing of high-order exceptional points \cite{chen2025quantum,li2025programmable,wu2026experimental}. Within this broad context, $\mathcal{PT}$-symmetric systems have emerged as fundamental and typical models \cite{Regensburger2012,Li2019,Ding2021,wang2021observation,quinn2023observing}. Several theoretical studies \cite{Deffner2015,Gardas2016,Wei2018,Zhou2021} have explored the validity of the Jarzynski equality under $\mathcal{PT}$-symmetric dynamics, but typically define energy using the full non-Hermitian Hamiltonian, resulting in complex spectra and limiting validity to the unbroken $\mathcal{PT}$ regimes. Recent progress shows that, by instead defining energy from the Hermitian part to keep the spectrum real and analyzing work statistics via postselected no-jump trajectories with symmetric transition probabilities, the non-Hermitian Jarzynski equality can be extended to both $\mathcal{PT}$-symmetric and $\mathcal{PT}$ broken regimes \cite{Erdamar2024}. Motivated by this finding, we pose the following question: In the framework of postselection, is $\mathcal{PT}$ symmetry a unique prerequisite for the validity of the non-Hermitian Jarzynski equality? Or, do broader classes of non-Hermitian systems exhibit similar thermodynamic properties?
\par In this work, we identify parity-exchange symmetry of transition probabilities as  the operative criterion  within a constructed two-level family of hybrid non-Hermitian Hamiltonians. Specifically, we study a $\mathcal{PT}$-$\mathcal{APT}$ family connected by $\mathrm{SU}(2)$ rotations and show that it belongs to a single $\mathrm{SU}(2)$-rotated orbit on which the conditional non-Hermitian Jarzynski equality is satisfied. Relative to Ref. \cite{Erdamar2024}, our contribution is therefore a criterion extension from the $\mathcal{PT}$ endpoint to a $\mathcal{PT}$-and-$\mathcal{APT}$ hybrid family. Experimentally, we implement three representative angles, $\theta_k \in \{0, \pi/4, \pi/2\}$, using a single trapped $^{171}\text{Yb}^{+}$ ion under static and time-dependent driving protocols. Because the implemented protocols are cyclic in the Hermitian spectrum, the measured relation reduces to the special case $\langle e^{-\beta W} \rangle = 1$. The scope of our conclusions is correspondingly restricted to two-level, no-jump, postselected dynamics with TPM energies defined by the Hermitian part.

\section{Theory}
\par For an open quantum system, the dynamics are in general governed by the Lindblad master equation. When we restrict our attention to trajectories that remain inside a given subspace and neglect quantum jumps from this subspace to the environment, the dynamics within the subspace can be described by an effective non-Hermitian Hamiltonian $H_{\mathrm{eff}}$. Importantly, this postselection method defines an experimentally implementable conditioned sub-ensemble (no-jump trajectories), rather than a purely mathematical construction.

\par Based on this method, we can explore the fluctuation relations for work in the presence of non-Hermitian dynamics. For the measurement of work, we employ the aforementioned TPM scheme and define the energy basis in terms of the eigenstates of the Hermitian part of the Hamiltonian \cite{Erdamar2024}. This framework resolves the ambiguity associated with complex energies of $H_{\rm eff}$, so that energy remains a genuine observable with a real spectrum and the work statistics admit a clear operational meaning (TPM scheme). Consequently, in a single realization the work is determined by the two Hermitian energy eigenvalues obtained in the initial and final projective measurements, while the whole non-Hermitian Hamiltonian affects the resulting distribution of work. We assume that the system is initially prepared in a thermal (Gibbs) state,
$\rho=\sum_{i=\pm} P_i\,|e_i\rangle\langle e_i|$,
where $|e_\pm\rangle$ are the eigenstates of the Hermitian part of the Hamiltonian with eigenvalues $\pm J_i$. The corresponding Boltzmann weights are $P_\pm = e^{\mp \beta J_i}/Z_i$, with the partition function $Z_i=e^{\beta J_i}+e^{-\beta J_i}$.
In the TPM protocol, a projective energy measurement at $t=0$ yields $E_i=\pm J_i$, the system evolves under the effective non-Hermitian Hamiltonian $H_{\mathrm{eff}}$ along a no-jump trajectory with propagator $K(T)=e^{-iH_{\text{eff}}T}$, and a second projective measurement at $t=T$ yields $E_f=\pm J_f$. The unnormalized transition probabilities are
\begin{equation}
    p_{fi} = \big|\langle e_f|K(T)|e_i\rangle\big|^2,
    \label{eq:unnormal}
\end{equation}
and we define the normalized conditional probabilities
\begin{equation}
    P_{fi} = \frac{p_{fi}}{\sum_f p_{fi}} = \frac{p_{fi}}{S_i(T)},\qquad \sum_f P_{fi}=1.
    \label{eq:normal}
\end{equation}
Here, the denominator $S_i(T) = \langle e_i|K^\dagger(T)K(T)|e_i\rangle = \sum_f p_{fi}$ represents the no-jump survival probability, which explicitly quantifies the theoretical postselection cost for each initial eigenstate. The joint probability for the process $E_i\to E_f$ is $P_{fi}P_i$, and the work is $W=E_f-E_i$. A straightforward calculation then gives
\begin{equation}
\begin{aligned}
    \big\langle e^{-\beta W}\big\rangle
    &= \sum_{i,f} e^{-\beta(E_f-E_i)}P_{fi}P_i \\
    &= \frac{1}{Z_i}\Big[
        e^{-\beta J_f}\big(1+P_{++}-P_{--}\big)
      + e^{\beta J_f}\big(1\\
      &\quad+P_{-+}-P_{+-}\big)
    \Big].
    \label{eq:cal}
\end{aligned}
\end{equation}
Comparing this expression with Eq.~\eqref{eq:JE}, we find that the equality holds identically if
\begin{equation}
    P_{++}=P_{--},\qquad P_{+-}=P_{-+}.
    \label{eq:con}
\end{equation}
This indicates that the conditional non-Hermitian Jarzynski equality is governed by a parity-exchange symmetry of the four transition probabilities. Here, the parity operator is understood in a generalized sense as swapping the two energy eigenstates within the two-level subspace. To mathematically formalize this concept, we introduce the generalized parity-exchange operator:
\begin{equation}
    \mathcal{P}_{\mathrm{ex}} = |e_+\rangle\langle e_-| + |e_-\rangle\langle e_+|.
    \label{parity}
\end{equation}
This abstract formulation explicitly acts as $\mathcal{P}_{\mathrm{ex}}|e_\pm\rangle = |e_\mp\rangle$, distinguishing it from standard spatial or spin parity operations. The relations in Eq.~\eqref{eq:con} essentially demand a macroscopic probability symmetry under the action of this specific basis exchange.
\par
\begin{figure}[H]
    \centering
    \includegraphics[width=8cm]{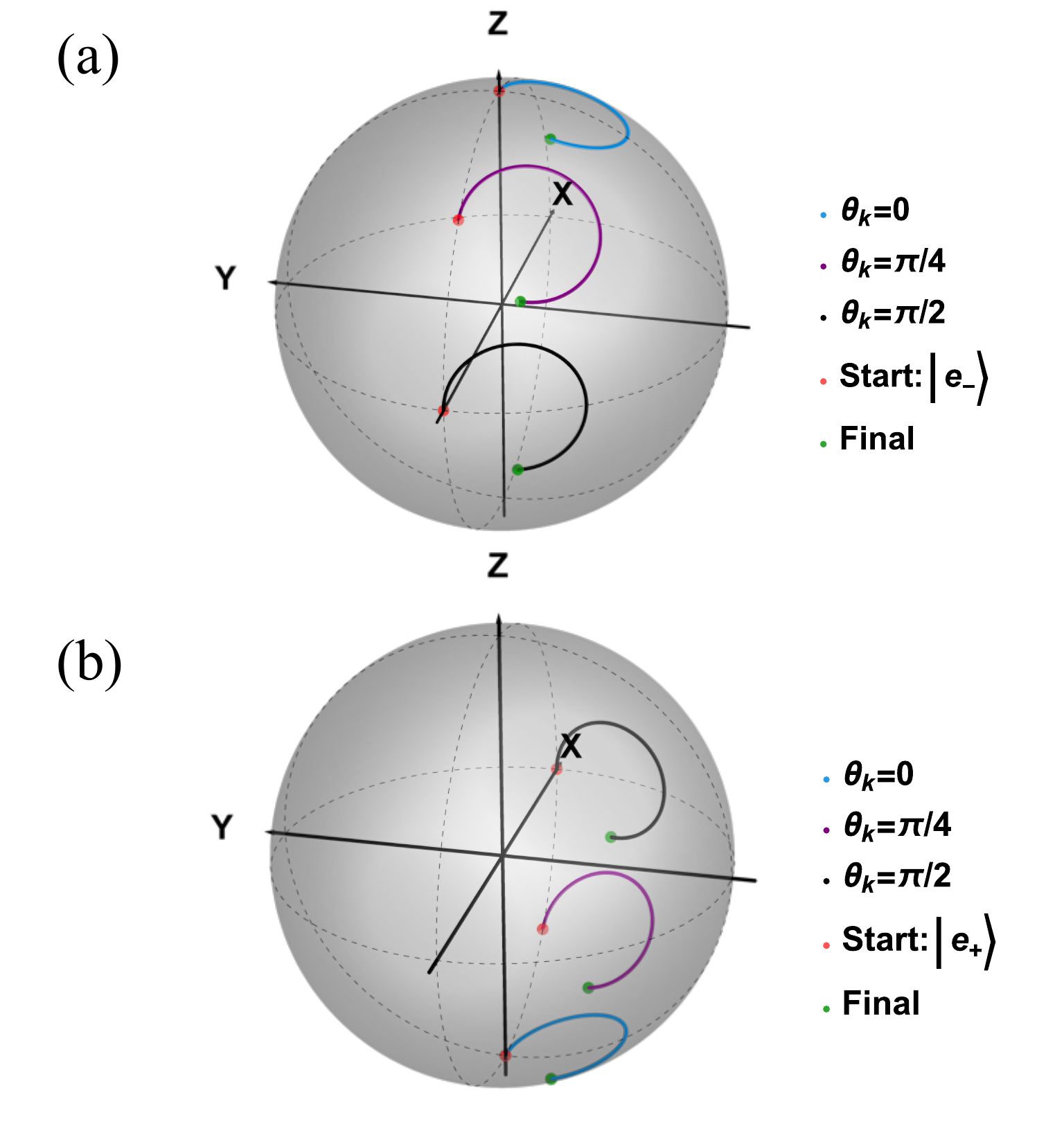}
    \caption{Evolution trajectories for quantum states $|e_-(\theta_k)\rangle$ and $|e_+(\theta_k)\rangle$ under different values of $\theta_k$. (a) Trajectories initialized at $|e_-(\theta_k)\rangle$. (b) Trajectories initialized at $|e_+(\theta_k)\rangle$. Different colors correspond to different $\theta_k$ values: blue ($\theta_k=0$), purple ($\theta_k=\pi/4$) and black ($\theta_k=\pi/2$). Initial and final states are marked in red and green, respectively.}
    \label{fig:multi}
\end{figure}

 Given this condition, we now focus on specific types of non-Hermitian systems, particularly two-level systems. Two typical classes of non-Hermitian Hamiltonians are the $\mathcal{PT}$-symmetric Hamiltonian and the $\mathcal{APT}$-symmetric Hamiltonian, which can be written as:
\begin{eqnarray}
    H_{\mathcal{PT}}=i\gamma\sigma_z+J\sigma_x,\quad
    H_{\mathcal{APT}}=i\gamma\sigma_x-J\sigma_z.
    \label{eq:PT}
\end{eqnarray}
Here, $J$ represents the coupling strength, while $\gamma$ denotes the strength of the non-Hermitian effect. The $\mathcal{PT}$-symmetric Hamiltonian satisfies the commutation relation $[\mathcal{PT}, H_{\mathcal{PT}}] = 0$, with $\mathcal{P}$ specified as $\sigma_x$ and $\mathcal{T}$ denoting complex conjugation, while the $\mathcal{APT}$-symmetric Hamiltonian obeys the anticommutation relation $\{\mathcal{PT}, H_{\mathcal{APT}}\} = 0$.
We now introduce a class of Hamiltonians obtained from these two prototypes:
\begin{equation}
\begin{aligned}
H_{\text{hb}}(\theta_k)&=\sin{\theta_k}H_{\mathcal{PT}}+\cos{\theta_k}H_{\mathcal{APT}}\\
          &=J(\sin{\theta_k}\sigma_x-\cos{\theta_k}\sigma_z)+i\gamma(\cos{\theta_k}\sigma_x\\
         &\quad+\sin{\theta_k}\sigma_z)\\
          &=H_{\text{HM}}(\theta_k)+H_{\text{NH}}(\theta_k),
\end{aligned}
\label{eq:ham}
\end{equation}
where $\theta_k$ parameterizes the hybridization, $H_{\text{HM}}(\theta_k)$ and $H_{\text{NH}}(\theta_k)$ denote the Hermitian and non-Hermitian components of the hybrid $\mathcal{PT}$-$\mathcal{APT}$ Hamiltonian $H_{\text{hb}}$, respectively. In our framework, the Hermitian part $H_{\text{HM}}(\theta_k)$ is taken as the system's energy operator \cite{Erdamar2024,Naghiloo2020}. It has two eigenstates:
\begin{equation}
    |e_-(\theta_k)\rangle=\begin{pmatrix}
      -\cos{\frac{\theta_k}{2}}\\
      \sin{\frac{\theta_k}{2}}
    \end{pmatrix},\qquad
    |e_+(\theta_k)\rangle=\begin{pmatrix}
      \sin{\frac{\theta_k}{2}}\\
      \cos{\frac{\theta_k}{2}}
    \end{pmatrix},
    \label{eq:e}
\end{equation}
which correspond to eigenenergies $-J$ and $J$, respectively. Importantly, the hybrid family $\{H_{\mathrm{hb}}(\theta_k)\}$ admits a natural SU(2) structure: by introducing the SU(2) rotation 
$U(\theta_k)=e^{-i(-\theta_k)\sigma_y/2}\in \text{SU}(2)$ we find that,
\begin{equation}
   H_{\mathrm{hb}}(\theta_k)=U(\theta_k)H_{\mathcal{APT}}U^{\dagger}(\theta_k).
   \label{eq:su}
\end{equation}
Varying $\theta_k$ continuously from $0$ to $2\pi$, the set $\{H_{\mathrm{hb}}(\theta_k)\}$ traces a closed $S^1$ curve in the operator space of $2\times2$ matrices, that is, a one-dimensional $S^1$ submanifold of the SU(2) adjoint orbit of $H_{\mathcal{APT}}$. On this $S^1$, the points $\theta_k=0$ and $\theta_k=\pi/2$ correspond to the $\mathcal{APT}$- and $\mathcal{PT}$-symmetric Hamiltonians, respectively. Here $U(\theta_k)$ acts as a global basis rotation on the two-dimensional Hilbert space. 
We now turn to the Jarzynski equality in the context of the above-mentioned non-Hermitian systems. To gain physical insight into the transition probabilities, we consider the normalized postselected state 
and plot the corresponding Bloch vector on the Bloch sphere. For different values of $\theta_k$, we plot the evolution trajectories starting from $|e_-(\theta_k)\rangle$ and $|e_+(\theta_k)\rangle$ defined by the Hermitian part $H_{\mathrm{HM}}(\theta_k)$, as shown in Fig.~\ref{fig:multi}(a) and Fig.~\ref{fig:multi}(b), respectively.
The evolution trajectories for various $\theta_k$ (representing $\mathcal{APT}$, hybrid, and $\mathcal{PT}$ dynamics) are plotted in Fig.~\ref{fig:multi}. For each fixed $\theta_k$, the trajectories initialized at $|e_-(\theta_k)\rangle$ and $|e_+(\theta_k)\rangle$ exhibit a striking mirror symmetry about the $y$-axis on the Bloch sphere. To formalize the physical implication of this visual symmetry, we describe the system state by a real Bloch vector $\mathbf{n}_{\pm}(T)$, where the subscript $\pm$ denotes the trajectory initialized from the energy eigenstate $|e_{\pm}\rangle$. Since a projective energy measurement amounts to projecting the postselected state onto the axis defined by the Hermitian internal energy operator $H_{\mathrm{HM}}(\theta_k)$, we define the measurement direction by the unit vector $\hat{\mathbf{n}}_{\mathrm{HM}}$. The transition probabilities are then geometrically defined as:
\begin{equation}
    \begin{aligned}
        P_{++} &= \frac{1}{2} \left( 1 + \mathbf{n}_{+}(T) \cdot \hat{\mathbf{n}}_{\mathrm{HM}} \right), \\
        P_{--} &= \frac{1}{2} \left( 1 - \mathbf{n}_{-}(T) \cdot \hat{\mathbf{n}}_{\mathrm{HM}} \right).
        \label{eq:n}
    \end{aligned}
\end{equation}
If this observed mirror symmetry translates to a  geometric constraint enforcing equal and opposite projections along the measurement axis ($\mathbf{n}_{+}(T) \cdot \hat{\mathbf{n}}_{\mathrm{HM}} = -\mathbf{n}_{-}(T) \cdot \hat{\mathbf{n}}_{\mathrm{HM}}$), substituting this relation into the geometric definitions in Eq.~\eqref{eq:n} naturally yields $P_{++} = P_{--}$. Furthermore, probability normalization inherently dictates that $P_{+-} = P_{-+}$, which constitutes the parity-exchange symmetry of the probabilities defined in Eq.~\eqref{eq:con}. To confirm that this geometric relationship reflects the operator structure of the dynamics rather than a visual coincidence, we now investigate the transition dynamics from an operator-level perspective.

At the core of this dynamic symmetry lies the parity-exchange operator defined in Eq.~\eqref{parity}, which takes the explicit angle-dependent form $\mathcal{P}_{\mathrm{ex}} = -\sin\theta_k \sigma_z - \cos\theta_k \sigma_x$. The Hamiltonian defined in Eq.~\eqref{eq:ham} then satisfies the algebraic symmetry:
\begin{equation}
    \mathcal{P}_{\mathrm{ex}} H^*_{\mathrm{hb}}(\theta_k) \mathcal{P}_{\mathrm{ex}} = -H_{\mathrm{hb}}(\theta_k).
\end{equation}
When exponentiated to the time-evolution operator $K(\theta_k,T) = \exp(-iH_{\mathrm{hb}}(\theta_k)T)$, and noting that the parity-exchange operator $\mathcal{P}_{\mathrm{ex}}$ is an involution, this Hamiltonian symmetry implies the relation:
\begin{equation}
    \mathcal{P}_{\mathrm{ex}} K^*(\theta_k,T) \mathcal{P}_{\mathrm{ex}} = K(\theta_k,T).
    \label{eq:symmetry}
\end{equation}
To see how this operator-level symmetry constrains the transition probabilities, we evaluate the transition amplitudes directly using Dirac notation. By applying the symmetry relation in Eq.~\eqref{eq:symmetry} and utilizing the swapping property $\mathcal{P}_{\mathrm{ex}} |e_\pm\rangle = |e_\mp\rangle$, we deduce the constraints on the survival amplitudes:
\begin{equation}
\begin{aligned}
    K_{++} &= \langle e_+ | K(\theta_k,T) | e_+ \rangle = \langle e_+ | \mathcal{P}_{\mathrm{ex}} K^*(\theta_k,T) \mathcal{P}_{\mathrm{ex}} | e_+ \rangle \\
    &= \langle e_- | K^*(\theta_k,T) | e_- \rangle = K_{--}^*.
\end{aligned}
\end{equation}
Similarly, for the cross-transition amplitudes:
\begin{equation}
\begin{aligned}
    K_{+-} &= \langle e_+ | K(\theta_k,T) | e_- \rangle = \langle e_+ | \mathcal{P}_{\mathrm{ex}} K^*(\theta_k,T) \mathcal{P}_{\mathrm{ex}} | e_- \rangle \\
    &= \langle e_- | K^*(\theta_k,T) | e_+ \rangle = K_{-+}^*.
\end{aligned}
\end{equation}
Taking the squared modulus
of these amplitude relations directly implies the absolute equality of the unnormalized probabilities:
\begin{equation}
    p_{++} = p_{--},\qquad p_{+-} = p_{-+}.
    \label{eq:unnormal1}
\end{equation} 
Because these unnormalized probabilities are symmetric, their corresponding survival factors must be  identical, $S_+(T) = p_{++} + p_{-+} = p_{--} + p_{+-} = S_-(T)$. Therefore, $P_{++}=p_{++}/S_+(T)=p_{--}/S_-(T)=P_{--}$ and $P_{-+}=p_{-+}/S_+(T)=p_{+-}/S_-(T)=P_{+-}$, which are required by the thermodynamic constraints in Eq.~\eqref{eq:con}.
 
In summary, since all hybrid Hamiltonians $H_{\mathrm{hb}}(\theta_k)$ are related by the SU(2) rotations $U(\theta_k)$ in Eq.~\eqref{eq:su} and lie on a connected $S^1$ orbit in operator space, the underlying operator-level symmetry $\mathcal{P}_{\mathrm{ex}} K^*(T) \mathcal{P}_{\mathrm{ex}} = K(T)$ holds throughout this constructed two-level orbit. Consequently, both the geometric mirror symmetry on the Bloch sphere and the parity-exchange symmetry of the transition probabilities are properties of this constructed SU(2)-rotated family and support the validity of the conditional non-Hermitian Jarzynski equality for the TPM construction used here.

It is worth noting that if the Hamiltonian is time-dependent, we can map it to a time-independent Hamiltonian model through the following equation:
\begin{equation}
    U(T) = \mathcal{T} \exp\left[-i \int_0^T H(t) \, dt\right] = \exp(-i H_{\text{F}} T),
    \label{eq:floquet}
\end{equation}
where $T$ denotes the total evolution duration, $U(T)$ is the time-evolution operator, $\mathcal{T}$ is the time-ordering operator, $H(t)$ represents the time-dependent Hamiltonian, and $H_{\text{F}}$ stands for the time-independent Floquet Hamiltonian (Appendix \ref{app:floquet}).

Finally, before detailing the experimental implementation, we briefly revisit our operational choice of the energy observable $H_{\mathrm{HM}}$ [Eq.~\eqref{eq:ham}] in the broader context of non-Hermitian thermodynamics. While multiple competing conventions exist (e.g., biorthogonal metrics or complex spectra), our specific choice of the Hermitian part ensures a real spectrum and enables a clear interpretation of work via the  TPM scheme. Consequently, the fluctuation relation tested in our experiment specifically concerns work defined with respect to $H_{\mathrm{HM}}$ under non-unitary no-jump evolution. We emphasize that the parity-exchange criterion in Eq.~\eqref{eq:con} is tied to this $H_{\mathrm{HM}}$-TPM construction; it is not invariant to alternative energy definitions, as non-orthogonal measurement bases would generally break the underlying geometric symmetry of the state projections. With this theoretical boundary established, we next present our experimental realization.

\section{Experimental setup} 
\begin{figure}[H]
    \centering
    \includegraphics[width=8cm]{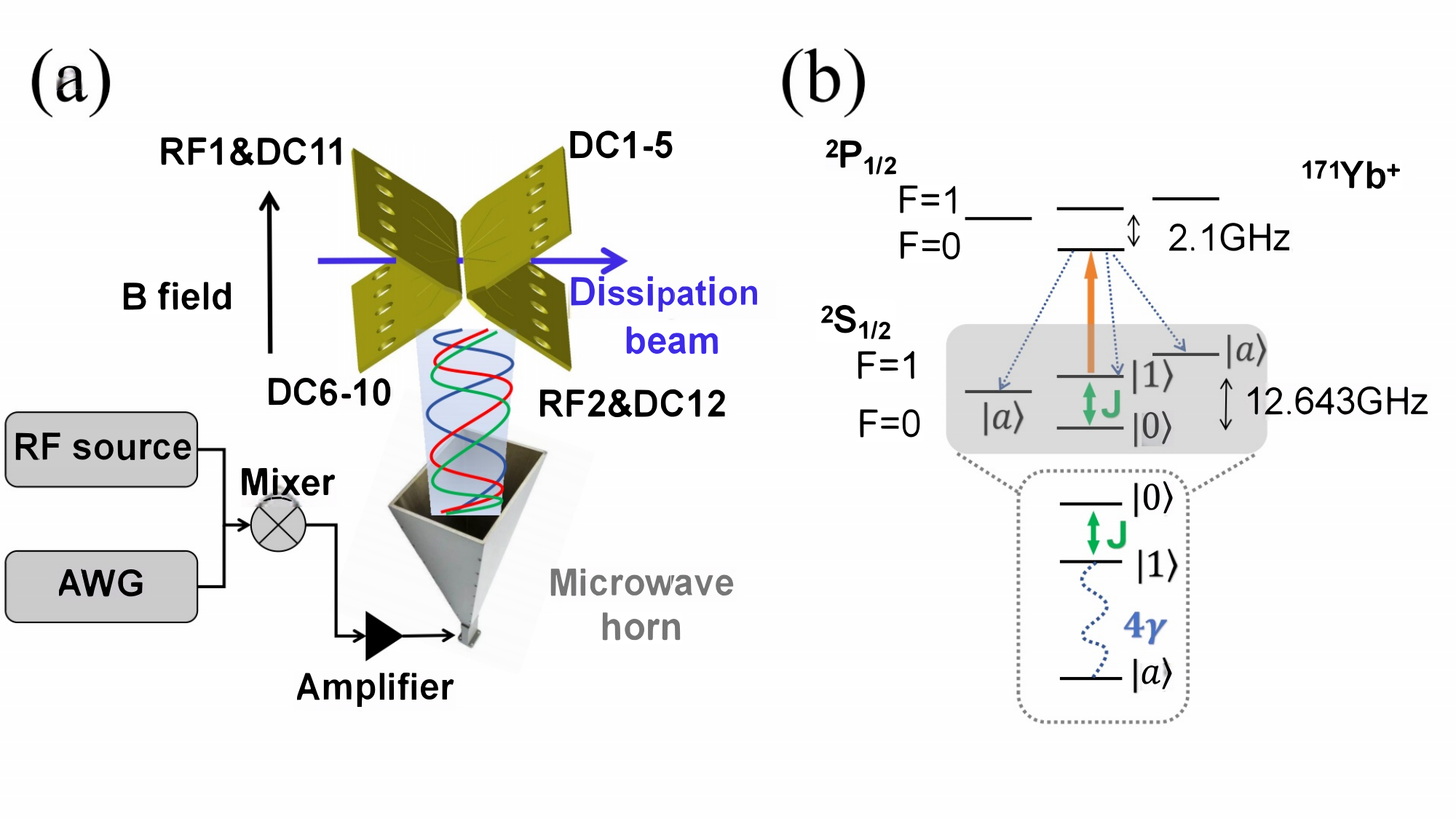}
    \caption{Experimental setup for verifying the Jarzynski equality. (a) Schematic of the trapped ion system, where the coherent driving terms $J(t)$ and $\Delta(t)$ are realized via applied microwave fields. (b) Implementation of the non-Hermitian Hamiltonian on a $^{171}$Yb$^+$ ion. The dissipation $\gamma$ is engineered by a 369.5 nm laser coupling the $|1\rangle$ state to the ${}^2P_{1/2}$ excited state, followed by spontaneous decay to the uncoupled auxiliary states $|a\rangle$. Conditional dynamics are extracted by postselecting only the no-jump trajectories remaining in the qubit subspace during the final readout.}
    \label{fig:trap}
\end{figure}

Our experiment is based on a $^{171}$Yb$^+$ ion confined at the center of a Paul trap, as shown in Fig.~\ref{fig:trap}(a), surrounded by four gold-plated ceramic blade electrodes: two RF electrodes and two DC electrodes, which provide RF signals and DC voltages to confine the ion in the trap. A magnetic field is applied along the Z-axis. The microwave signal used to drive the qubit is generated by mixing a 12.611 GHz signal from a standard signal generator with a 31.25 MHz signal from an arbitrary waveform generator (AWG), which enables coupling between the spin states $|0\rangle=|F=0,m_F=0\rangle$ and $|1\rangle=|F=1,m_F=0\rangle$. Additionally, a dissipation beam with a wavelength of 369.5 nm is applied along the Y-direction (with the trap axis along the X-direction). This beam drives the transition from $|1\rangle$ to the $^2P_{1/2}$ excited state, followed by spontaneous emission to the three magnetic levels $|F=1,m_F=0,\pm1\rangle$, where decay to the states $|F=1,m_F=\pm1\rangle$ can be treated as a dissipative part.
\par Building upon early pioneering demonstrations of trapped-ion non-Hermitian physics \cite{wang2021observation,Ding2021}, we can construct a non-Hermitian evolution model using the experimental system described above \cite{Lu2024a,Lu2024b}. As shown in Fig.~\ref{fig:trap}(b), the multi-level evolution can be described by the Lindblad equation:
\begin{equation}
    \frac{d\rho(t)}{dt}=-i[H_c(t),\rho(t)]+(L_1\rho(t)L_1^{\dagger}-\frac{1}{2}\{L_1^{\dagger}L_1,\rho(t)\}),
\end{equation}
where $H_c(t)=J(|0\rangle\langle1|+|1\rangle\langle0|)+\Delta/2(|0\rangle\langle0|-|1\rangle\langle1|)$, $\Delta(t)$ is the detuning, and $L_1=\sqrt{4\gamma}|a\rangle\langle1|$ is the dissipation operator corresponding to the decay from level $|1\rangle$ to $|a\rangle=|F=1,m_F=\pm1\rangle$. 
Since level $|a\rangle$ is dynamically decoupled from the qubit, we can employ a postselection method \cite{Naghiloo2019,Lu2024b} to restrict our analysis to quantum trajectories within the qubit subspace. In this case, the total effective Hamiltonian for the two-level system is expressed as \cite{Lu2024b}:
\begin{eqnarray}
    H_{\text{total}}(t)=J(t)\sigma_x+\Delta(t)\sigma_z/2-2i\gamma|1\rangle\langle1|.
\end{eqnarray}
By algebraically decomposing the physical dissipation term as $-2i\gamma|1\rangle\langle 1| = i\gamma\sigma_z - i\gamma\mathbb{I}$, the system naturally manifests a relative non-Hermitian core $H_{\text{relative}}(t) = J(t)\sigma_x + [\Delta(t)/2 + i\gamma]\sigma_z$, and a global scalar decay $-i\gamma\mathbb{I}$. This global dissipative term exclusively dictates the experimental postselection cost (survival probability) and cancels out in the normalized conditional probabilities $P_{fi}$, leaving the conditional transition dynamics governed by $H_{\text{relative}}(t)$. 

To experimentally implement the generalized hybrid Hamiltonian $H_{\text{hb}}(\theta_k)$, we engineer a parameterized pulse sequence based on this driven-dissipative system: applying a pair of microwave rotation pulses with an angle $\phi_k = \theta_k - \pi/2$ along the $\pm y$ axis before and after the evolution under $H_{\text{total}}$. This global SU(2) transformation maps the time evolution operator to 
\begin{equation}
   U'(\theta_k, T) = \mathcal{T} \exp \left[ -i \int_0^T H'_{\text{total}}(\theta_k, t) \, dt \right],
    \label{eq:U'}
\end{equation}
where the rotated general Hamiltonian is explicitly given by
\begin{equation}
\begin{aligned}
    H'_{\text{total}}(\theta_k, t) = &J(t)(\sin\theta_k \sigma_x - \cos\theta_k \sigma_z) + [\Delta(t)/2 + i\gamma]\\
    &(\cos\theta_k \sigma_x + \sin\theta_k \sigma_z) - i\gamma\mathbb{I}.
\end{aligned}
\end{equation}
Notably, the global scalar decay term commutes with the rotations and remains invariant. When setting the detuning $\Delta(t) =0$, this transformed Hamiltonian exactly reduces to the target generalized model, simulating $H_{\text{hb}}(\theta_k)$. By tuning the experimental parameter $\theta_k$, this framework allows us to smoothly interpolate between the $\mathcal{PT}$-symmetric endpoint ($\theta_k = \pi/2$) and the  $\mathcal{APT}$ endpoint ($\theta_k = 0$) \cite{Bian2023}. In the present passive trapped-ion implementation, this interpolation is realized through basis rotations around the same dissipative core evolution. Thus the $\mathcal{APT}$ endpoint and the intermediate hybrid point should be understood as rotated-basis realizations of the
constructed family, rather than as independent active gain and loss implementations.
\par Based on the non-Hermitian TPM scheme, our experiments are executed in the following three consecutive steps:

(i) \textbf{Deterministic Eigenstate Preparation (Effective First Measurement)}. In a standard TPM protocol, an initial Gibbs state $\rho = e^{-\beta H_{\text{HM}}(\theta_k)}/Z$ undergoes a projective measurement, collapsing into one of the energy eigenstates $|e_{\pm}(\theta_k)\rangle$ with a Boltzmann probability $P_{\pm} = e^{\mp \beta J}/Z$. To reproduce these post-measurement statistics without performing a physical first measurement, we directly initialize the trapped ion into the pure eigenstates $|e_{+}(\theta_k)\rangle$ and $|e_{-}(\theta_k)\rangle$ in separate experimental runs and combine the outcomes with the corresponding Boltzmann weights. Consistent with the hybrid family defined in Section 2, the $\mathcal{PT}$, $\mathcal{APT}$, and intermediate hybrid processes correspond to hybridization angles $\theta_k=\pi/2$, $\theta_k=0$, and $\theta_k=\pi/4$, respectively. Consequently, the internal energy operator $H_{HM}(\theta_k)$ from Eq. (9) reduces to the specific measurement bases: $\sigma_x$ (for $\mathcal{PT}$), $\sigma_z$ (for $\mathcal{APT}$), and the rotated eigenbasis defined by Eq.~\eqref{eq:e} (for the $\theta_k=\pi/4$ case). This deterministic preparation is achieved by applying parameter-dependent rotation pulses to the standard $|0\rangle$ state. To confirm our experimental capability to prepare the actual thermal mixture, we also independently verified the Gibbs state initialization via an active random-phase dephasing protocol, as detailed in Appendix \ref{app:mixed_state}.

(ii) \textbf{Applying work on the qubit}. We then implement the non-Hermitian work process by applying the generalized microwave pulse sequence $U'(\theta_k, T)$ to the qubit, as derived in Eq.~\eqref{eq:U'}. The total evolution time $T$ is varied from $10\,\mu\text{s}$ to $50\,\mu\text{s}$ in steps of $2\,\mu\text{s}$, during which the core relative dynamics are driven by adjusting the coupling $J(t)$ and detuning $\Delta(t)$. Specifically, we adopt the following three temporal driving protocols, as Eqs.~\eqref{eq:const}--\eqref{eq:dchange}.
 \begin{figure*}[!htbp]
       \begin{gather}
    J(t) = J_1, \quad \Delta(t) = 0 \label{eq:const}  \\
    J(t) = 
    \left\{
    \begin{aligned}
        &J_{\text{min}} + (J_{\text{max}} - J_{\text{min}}) \frac{2t}{T}, &&0 \leqslant t < T/2 \\
        &J_{\text{max}} + (J_{\text{min}} - J_{\text{max}}) \frac{2(t - T/2)}{T}, && T/2 \leqslant t \leqslant T
    \end{aligned}
    \right. ,\quad \Delta(t) = 0 
    \label{eq:change}\\
    J(t) = J_2, \quad \Delta(t) = \Delta_1\sin(2\pi t/T). \label{eq:dchange}
    \end{gather} 
   \end{figure*}  

(iii) \textbf{Final measurement}. After the non-Hermitian evolution, the second projection measurement is performed. Crucially, the measurement basis follows the generalized energy eigenstates $\{|e_{+}(\theta_k)\rangle, |e_{-}(\theta_k)\rangle\}$ determined by the initial hybridization angle $\theta_k$. Due to the non-unitary nature of the evolution, the conditional probability of transitioning from an initial eigenstate $|i\rangle$ to a final eigenstate $|f\rangle$ is normalized by the survival probability \cite{Bian2023}:
\begin{eqnarray}
    P_{fi}(\theta_k, T) = \frac{|\langle f|U'(\theta_k, T)|i\rangle|^2}{\langle i|U'^{\dagger}(\theta_k, T)U'(\theta_k, T)|i\rangle}.
    \label{eq:pro}
\end{eqnarray}
Here, $i, f \in \{|e_{+}(\theta_k)\rangle, |e_{-}(\theta_k)\rangle\}$. As a result, for each sampled hybridization angle $\theta_k \in \{0, \pi/4, \pi/2\}$, we experimentally obtain four generalized conditional probabilities $\{P_{++}, P_{-+}, P_{+-}, P_{--}\}$ within its respective eigenbasis. These normalized probabilities are subsequently used to calculate the exponential work according to Eq.~(\ref{eq:cal}), allowing us to test the generalized Jarzynski equality on this representative set of hybrid points.

Practically, the effective non-Hermitian dynamics involves a global dissipative term $-i\gamma \mathbb{I}$, causing the survival probability to decay exponentially as $e^{-2\gamma t}$ (Appendix \ref{app:survival}). Although this scalar damping  cancels out in the normalized transition probabilities, the associated continuous population loss still suppresses the experimental detection efficiency. To maintain a sufficient signal-to-noise ratio, we implement a piecewise evolution scheme \cite{Lu2024a}. We divide the total evolution time $T$ into $N$ discrete segments. For the $n$-th segment ($1 \le n \le N$), the qubit is explicitly initialized in the intermediate state corresponding to $t_{n-1} = (n - 1)T/N$ based on theoretical predictions. The system is then subjected to physical non-Hermitian evolution governed by $H_{\mathrm{eff}}$ for a short duration $\Delta t = T/N$. This piecewise strategy alleviates the continuous exponential drop and allows us to map out the non-Hermitian trajectory without being dominated by the global loss overhead.

\section{Results and Discussion}

To evaluate the Jarzynski equality, we weight the measured transition probabilities $P_{fi}$ [Eq.~\eqref{eq:pro}] with the theoretical initial Gibbs distribution $P_i$ to calculate the exponential work. Because the protocols considered here are cyclic in the Hermitian spectrum, the tested relation reduces to the special case $\langle e^{-\beta W}\rangle = 1$. Experimentally, we set $\beta = 20 \,\mu\text{s/rad}$ and $\gamma = 0.02\,\mu\text{s}^{-1}$, with 4500 repetitions for each transition. The mixed-state preparation and tomography reported in Appendix~\ref{app:mixed_state} serve as an independent calibration showing that the target populations associated with the effective inverse temperature $\beta$ can be prepared with high fidelity. We emphasize that, in the Jarzynski equality evaluation, $\beta$ is used to assign the theoretical initial Gibbs weights rather than to represent coupling to a real thermal bath.

\par
\begin{figure}[H]
    \centering
    \includegraphics[width=8cm]{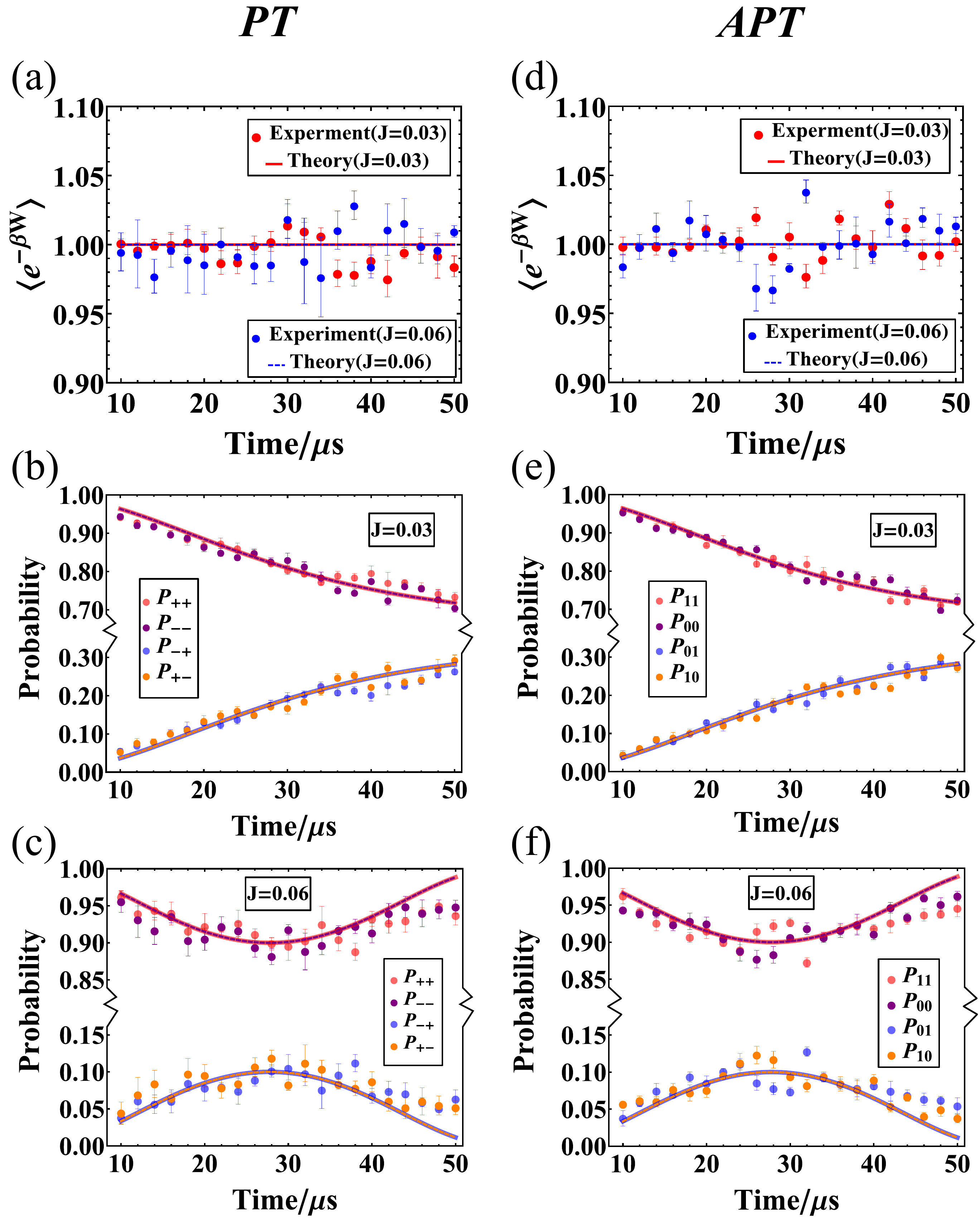}
       \caption{Experimental and theoretical results of the exponential work $\langle e^{-\beta W}\rangle$ and transition probabilities under $\mathcal{PT}$ and $\mathcal{APT}$ evolution with constant $J$. Panels (b) and (e) show the
       transition probabilities at $J = 0.03\,\mu\text{s}^{-1}$, while panels (c) and (f) show the corresponding results at $J = 0.06\,\mu\text{s}^{-1}$. The data points with error bars represent experimental results, while the solid or dashed lines denote theoretical results. The vertical axes in (b), (c), (e) and (f) are broken to make the experimental error bars clearly visible. }
       \label{fig:Jconst}
\end{figure}

 We first set the detuning to zero and keep $J$ constant over time, according to Eq.~(\ref{eq:const}), measuring the distribution of work after evolution under $\mathcal{PT}$ and $\mathcal{APT}$ symmetries, as shown in Fig.~\ref{fig:Jconst}. Panels~\ref{fig:Jconst}(a)--~\ref{fig:Jconst}(c) correspond to $\mathcal{PT}$ evolution, whereas Fig.~\ref{fig:Jconst}(d)--~\ref{fig:Jconst}(f) correspond to $\mathcal{APT}$ evolution. Fig.~\ref{fig:Jconst}(a) and Fig.~\ref{fig:Jconst}(d) show the results of exponential work calculations when $J$ is set to $J = 0.03\,\mu\text{s}^{-1}$ (red) and $J = 0.06\,\mu\text{s}^{-1}$ (blue). Fig.~\ref{fig:Jconst}(b) and Fig.~\ref{fig:Jconst}(e) display the transition probabilities for $\mathcal{PT}$ and $\mathcal{APT}$ evolution at $J = 0.03\,\mu\text{s}^{-1}$, while Fig.~\ref{fig:Jconst}(c) and Fig.~\ref{fig:Jconst}(f) show the transition probabilities for $\mathcal{PT}$ and $\mathcal{APT}$ evolution at $J = 0.06\,\mu\text{s}^{-1}$. From Fig. \ref{fig:Jconst}(a) and \ref{fig:Jconst}(d), it can be seen that under constant $J$, both cases satisfy the relation $\langle e^{-\beta W}\rangle \approx 1$, indicating that the Jarzynski equality holds under these conditions. As predicted by the geometric analysis, the measured probabilities exhibit the expected parity-exchange symmetry both in $\mathcal{PT}$ and $\mathcal{APT}$ processes, as shown in Fig.~\ref{fig:Jconst}(b), Fig.~\ref{fig:Jconst}(c), Fig.~\ref{fig:Jconst}(e) and Fig.~\ref{fig:Jconst}(f). By comparing Fig.~\ref{fig:Jconst}(b) and \ref{fig:Jconst}(e), as well as Fig.~\ref{fig:Jconst}(c) and \ref{fig:Jconst}(f), we observe that under the same driving $J(t)$ settings, the set of probabilities $\{P_{++}, P_{--}\}$ obtained from $\mathcal{PT}$ evolution corresponds to $\{P_{11}, P_{00}\}$ in $\mathcal{APT}$ evolution, while $\{P_{-+}, P_{+-}\}$ corresponds to $\{P_{01}, P_{10}\}$. This correspondence arises because, in the present passive implementation, the $\mathcal{PT}$ and $\mathcal{APT}$ evolutions are realized as different $\theta_k$-dependent measurement and preparation bases within the same constructed SU(2)-rotated family.

\par
\begin{figure}[H]
    \centering
    \includegraphics[width=8cm]{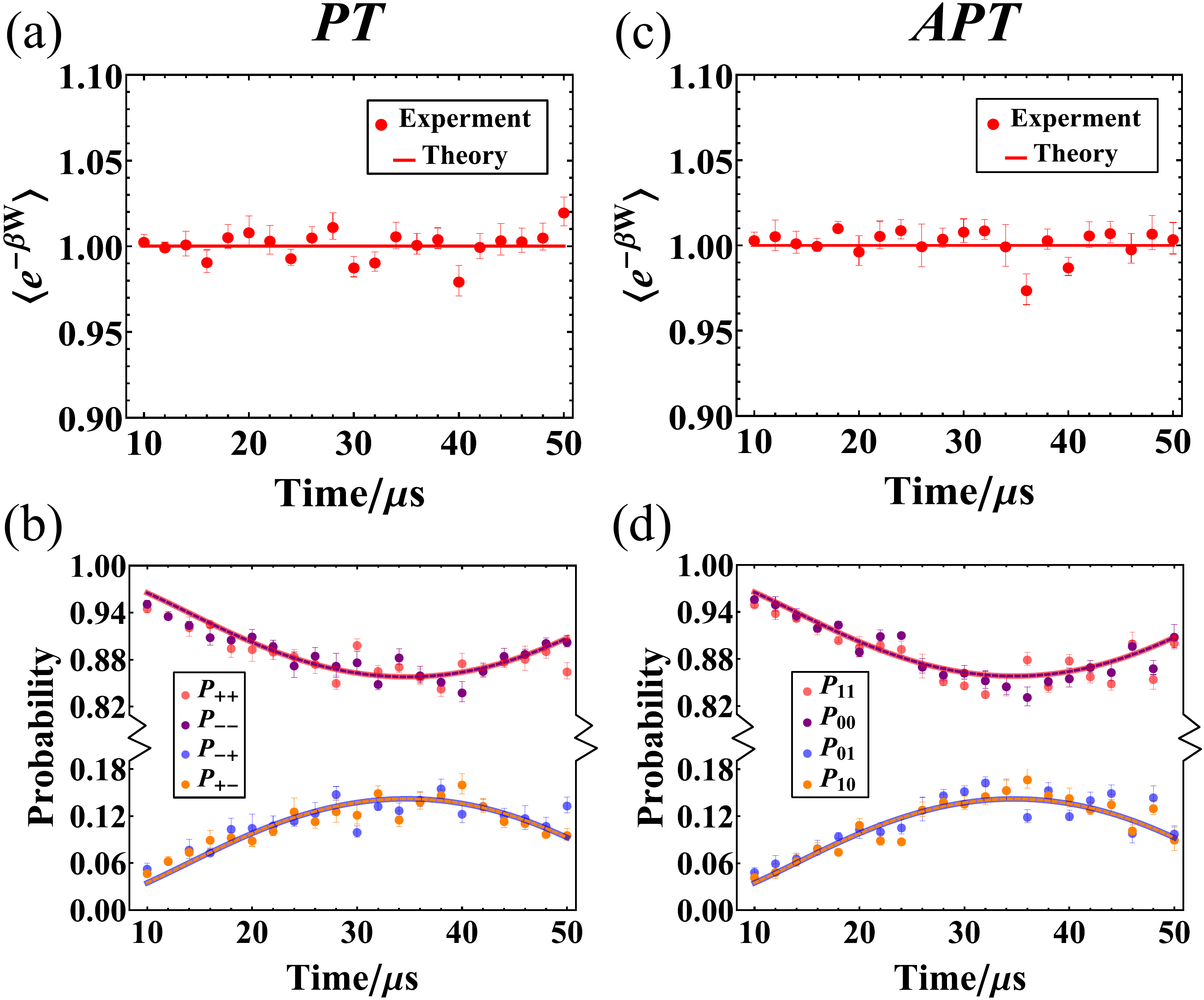}
    \caption{Experimental and theoretical results of the exponential work $\langle e^{-\beta W}\rangle$ and transition probabilities under $\mathcal{PT}$ and $\mathcal{APT}$ evolution, where $J$ varies linearly from $0.03\,\mu\text{s}^{-1}$ to $0.06\,\mu\text{s}^{-1}$ and then returns to $0.03\,\mu\text{s}^{-1}$. The data points with error bars represent experimental results, while the solid or dashed lines denote theoretical results. The vertical axes in (b) and (d) are broken to make the experimental error bars clearly visible. (a) and (b) correspond to the exponential work and transition probabilities under $\mathcal{PT}$ evolution, respectively, while (c) and (d) correspond to the exponential work and transition probabilities under $\mathcal{APT}$ evolution, respectively.}
    \label{fig:Jchange}
\end{figure}

 Next, maintaining zero detuning, we vary $J(t)$ linearly from $ 0.03\,\mu\text{s}^{-1}$ to $0.06\,\mu\text{s}^{-1}$ and then return to $0.03\,\mu\text{s}^{-1}$, according to Eq.~(\ref{eq:change}). The results of the exponential work and transition probabilities are shown in Fig.~\ref{fig:Jchange}. From Fig.~\ref{fig:Jchange}(a) and \ref{fig:Jchange}(c), it can be observed that the relation $\langle e^{-\beta W}\rangle \approx 1$ is also satisfied under these conditions. Furthermore, from Fig.~\ref{fig:Jchange}(b) and \ref{fig:Jchange}(d), it can be seen that the transition probabilities for $\mathcal{PT}$ and $\mathcal{APT}$ evolution exhibit symmetries similar to those obtained under constant $J$, as well as a one-to-one correspondence between the two cases.

\begin{figure}[H]
    \centering
    \includegraphics[width=8cm]{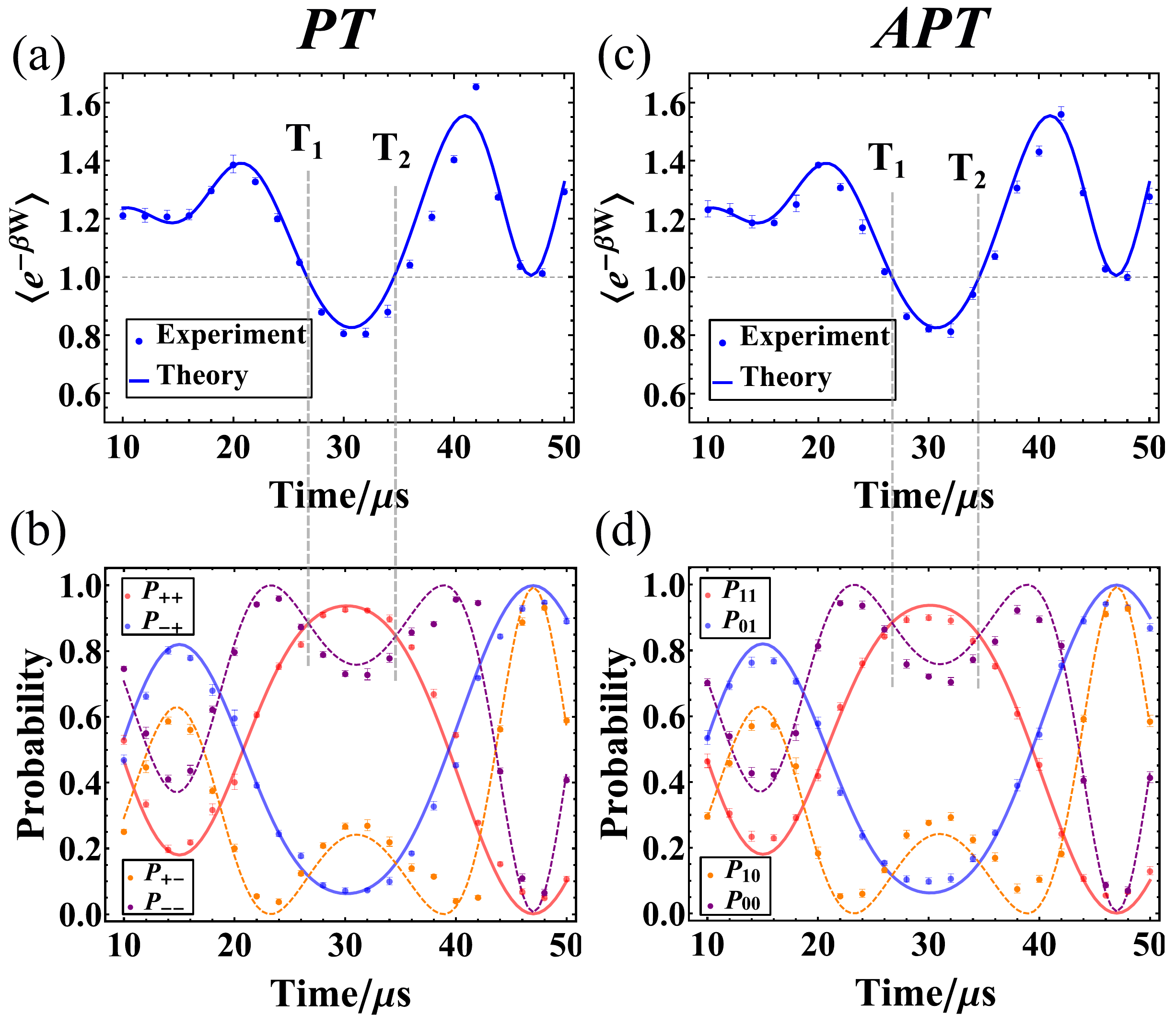}
    \caption{The theoretical and experimental results of the exponential work $\langle e^{-\beta W}\rangle$ and transition probabilities with detuning added to the $\mathcal{PT}$ and $\mathcal{APT}$ Hamiltonians. (a) and (c) show the cases where a sinusoidal modulation of detuning $\Delta(t) = 0.5\sin\frac{2\pi t}{T}\,\mu\text{s}^{-1}$ is introduced into the $\mathcal{PT}$ and $\mathcal{APT}$ Hamiltonians, respectively. The points with error bars represent the experimental results, while the solid lines correspond to the theoretical results. (b) and (d) correspond to the transition probabilities under evolution for a $\mathcal{PT}$-symmetric Hamiltonian with detuning and an $\mathcal{APT}$-symmetric Hamiltonian with detuning, respectively. $T_1$ and $T_2$ denote the evolution times at which the exponential work equals 1.}
    \label{fig:Dwork}
\end{figure}

Finally, we introduce detuning into both the $\mathcal{PT}$ and $\mathcal{APT}$ Hamiltonians to examine whether the relationship $\langle e^{-\beta W}\rangle = 1$ still holds. As shown in Eq.~(\ref{eq:dchange}), we introduce a sinusoidally modulated detuning that completes exactly one oscillation period, where the time integral of the detuning over the total duration equals zero. In the detuning experiments, $J_2 = 0.12\,\mu\text{s}^{-1}$ and $\Delta_1=0.5\,\mu\text{s}^{-1}$. Fig.~\ref{fig:Dwork} presents the experimental and theoretical results for the exponential work. We can see that after introducing detuning into the $\mathcal{PT}$-symmetric and $\mathcal{APT}$-symmetric Hamiltonians, the transition probabilities lose their symmetry for most of the evolution time, and the exponential work does not satisfy the relation $\langle e^{-\beta W}\rangle = 1$, which indicates the violation of the Jarzynski equality. Only at specific evolution times $T_1 \approx 26.7\,\mu\text{s}$ and $T_2 \approx 34.6\,\mu\text{s}$, the transition probabilities regain the parity-exchange symmetry, as shown in Fig.~\ref{fig:Dwork}(b) and \ref{fig:Dwork}(d), where the extracted Floquet Hamiltonians recover the hybrid form defined in Eq.~\eqref{eq:ham}, which can be theoretically calculated in Appendix \ref{app:floquet}. At these instances, the exponential work satisfies the relation $\langle e^{-\beta W}\rangle \approx 1$, as indicated by the intersections of the dashed reference line with the measured and theoretical curves in Fig. \ref{fig:Dwork}(a) and Fig. \ref{fig:Dwork}(c). Conceptually, introducing detuning $\Delta(t)$ shifts the instantaneous Hamiltonian outside the protected $\mathrm{SU}(2)$ subspace [Eq. (9)]. Consequently, the evolution operator $K(T)$ generally violates the algebraic symmetry $\mathcal{P}_{ex}K^*(T)\mathcal{P}_{ex} = K(T)$ [Eq.~\eqref{eq:symmetry}], breaking the parity-exchange probabilities. However, at specific evolution times $T_1 \approx 26.7~\mu\text{s}$ and $T_2 \approx 34.6~\mu\text{s}$, the dynamically accumulated symmetry-breaking components within the effective Floquet Hamiltonian $H_F(T)$ evaluate to zero. At these instances, $K(T)$ algebraically returns to the canonical hybrid form, temporarily restoring the parity-exchange symmetry. To further probe these symmetry revivals, we performed high-resolution temporal scans ($0.2\,\mu\text{s}$ steps) around $T_1$ and $T_2$. As detailed in Appendix \ref{app:finescan} for a representative $\mathcal{PT}$-symmetric case, these measurements resolve the zero-crossing of the transition-probability difference and support the consistency of the piecewise protocol.

While the $\mathrm{SU}(2)$ rotation picture holds  at the theoretical operator level, experimentally evaluating an intermediate state provides a clearer demonstration of the hybrid family than testing only the two endpoint models. To this end, we performed full measurements at the maximally hybridized angle $\theta_k = \pi/4$ (theoretically visualized by the purple trajectories in Fig.~\ref{fig:multi}). As detailed in Appendix~\ref{app:pi4}, the experimental data are in good agreement with the theoretical predictions. This intermediate-point verification provides an additional experimental check that is consistent with the preservation of
the parity-exchange symmetry at a representative interior point of the implemented SU(2) orbit.
    
\section{Conclusion}
In summary, we have shown that, within the present two-level postselected TPM framework, the validity of the conditional Jarzynski equality in non-Hermitian quantum systems is governed by a parity-exchange symmetry of transition probabilities, rather than being exclusive to $\mathcal{PT}$ symmetry. Within a postselected no-quantum-jump framework, defining work through the Hermitian part of a hybrid $\mathcal{PT}$--$\mathcal{APT}$ Hamiltonian family reveals that mirror-related trajectories on the Bloch sphere give rise to these symmetric transition channels under the constructed operator-level symmetry. Experimentally, we implemented representative $\mathcal{PT}$- and $\mathcal{APT}$-symmetric Hamiltonians, alongside an intermediate hybrid point, using a single trapped ${}^{171}\text{Yb}^+$ ion, and measured the work distributions via a Hermitian-part TPM protocol under non-Hermitian evolution. Specifically, driving protocols without detuning restrict the Hamiltonian to the protected $\mathrm{SU}(2)$ subspace, preserving the algebraic symmetry $\mathcal{P}_{ex}K^*(T)\mathcal{P}_{ex} = K(T)$ and maintaining the Jarzynski equality. In contrast, time-dependent detuning shifts the system outside this subspace, violating the symmetry and breaking the equality, which is then restored only at specific instances when the effective Floquet Hamiltonian dynamically regains the parity-exchange symmetry.

Looking forward, the passive non-Hermitian framework involves a tradeoff between accessing exact conditional statistics and retaining detection efficiency under postselection-induced population decay. Although we mitigated part of this overhead in the present work by implementing a piecewise evolution scheme \cite{Lu2024a}, transitioning to an active $\mathcal{PT}$-symmetric platform with deterministic balanced gain and loss could reduce this bottleneck. Similar to recent protocols utilizing active $\mathcal{PT}$ symmetry to circumvent fidelity-entanglement tradeoffs \cite{liu2024using}, an active $\mathcal{PT}$-symmetric implementation could, in principle, reduce or remove the postselection overhead associated with passive no-jump evolution. However, the corresponding fluctuation relation would have to be re-examined in the presence of gain-induced noise and platform-dependent implementation details. Beyond trapped ions, related hybrid $\mathcal{PT}$--$\mathcal{APT}$ optical platforms, where nanoparticle perturbations enable control of spectral transitions and photonic transmission~\cite{Zhang2025PRA}, may provide another possible route for exploring symmetry-controlled non-Hermitian dynamics.

\par Our results extend fluctuation relations with a constructed class of non-Hermitian dynamics and clarify the central role of postselection-induced parity-exchange symmetry in non-equilibrium quantum thermodynamics. This symmetry-enforced perspective provides a route to engineering and testing non-Hermitian fluctuation theorems in open quantum platforms, and it paves the way for exploring higher-dimensional or more complex non-Hermitian systems  and their integration with quantum technologies operating far from equilibrium.
\section*{Acknowledgements}
This work was supported by the National Key Research and Development Program of China, the ``Gravitational Wave Detection'' Special Project under Grant No.~2022YFC2204402, the Guangdong Provincial Quantum Science Strategic Initiative under Grants No.~GDZX2203001 and GDZX2303003, the State Key Laboratory of Optoelectronic Materials and Technologies (Sun Yat-sen University) under Grant No.~OEMT-2025-KF-06, and the Shenzhen Science and Technology Program under Grant No.~JCYJ20220818102003006.
\appendix 
\setcounter{figure}{0}
\renewcommand{\thefigure}{A\arabic{figure}}
\titleformat{\section}[block]
  {\large\bfseries} 
  {Appendix \thesection}         
  {1em}                          
  {\noindent}                    
\section{Preparation and Tomography of the Initial Mixed State}
\label{app:mixed_state}

Testing the conditional Jarzynski relation requires a precise initial population distribution. Using the $\mathcal{APT}$ condition as an illustrative case, we prepare the required $\sigma_z$-basis thermal mixed state via an active random-phase dephasing protocol. This approach effectively suppresses quantum coherences while maintaining the diagonal populations dictated by the preset inverse temperature $\beta = 20\,\mu\text{s}/\text{rad}$ and $J = 0.03\,\mu\text{s}^{-1}$.

\begin{figure}[H]
    \centering
    \includegraphics[width=8cm]{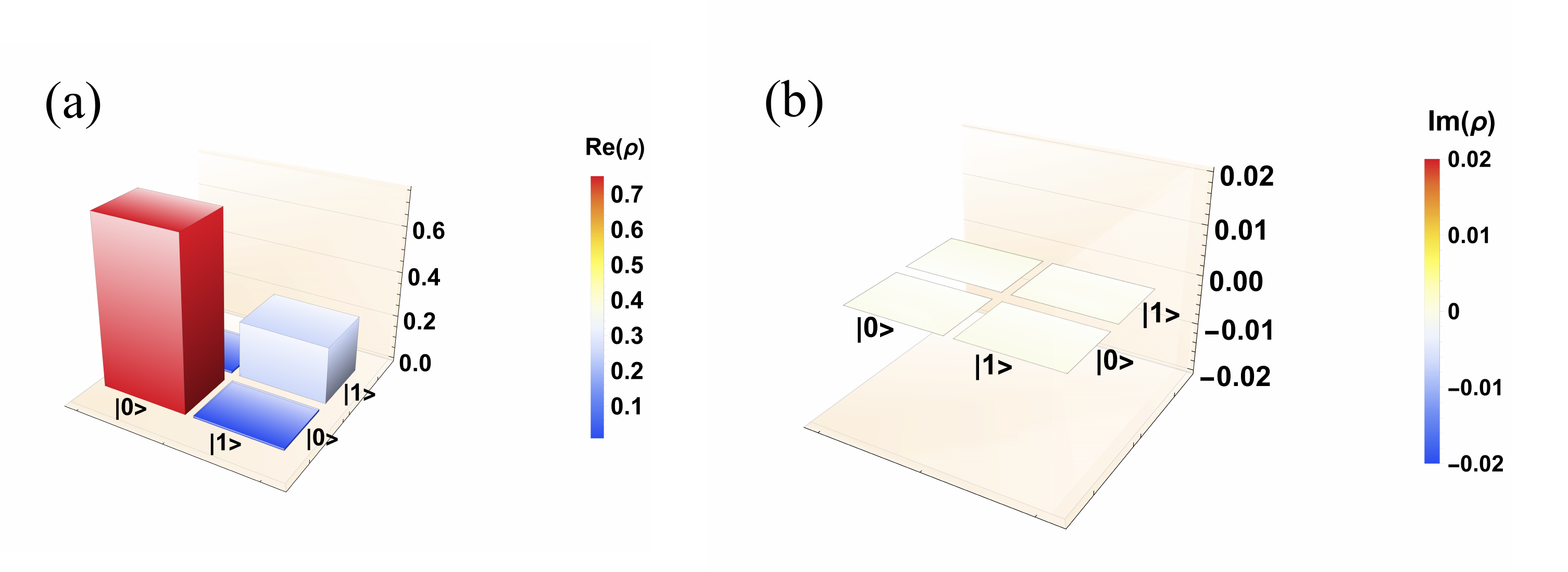}
    \caption{\textbf{Quantum state tomography of the initial mixed state.} (a) Real and (b) imaginary parts of the experimentally reconstructed density matrix $\rho_{\text{exp}}$ prepared in the $\sigma_z$ basis (corresponding to the $\mathcal{APT}$ condition). Determined by the effective inverse temperature $\beta = 20 \,\mu\text{s/rad}$ and $J=0.03\,\mu\text{s}^{-1}$, the theoretical target is diagonal: $\rho_{\text{theory}} = \mathrm{diag}(0.768, 0.232)$. Consistent with this theoretical expectation, the experimental state yields diagonal elements of $0.751 \pm 0.014$ and $0.252 \pm 0.014$. The off-diagonal real components ($0.009 \pm 0.015$) and the imaginary components are consistent with zero within experimental uncertainty, demonstrating the effectiveness of the active random-phase dephasing protocol. The error margins represent the shot noise propagated from 4500 experimental repetitions.}
    \label{fig:mixed_state}
\end{figure}

To quantify the fidelity of this state initialization, we performed full quantum state tomography on the prepared mixed state immediately following the dephasing protocol. Fig.~\ref{fig:mixed_state} presents the real and imaginary parts of the experimentally reconstructed density matrix, $\rho_{\text{exp}}$. 
Consistent with the theoretical target $\rho_{\text{theory}} = \text{diag}(0.768, 0.232)$, the experimental density matrix yields diagonal populations of $\rho_{00} = 0.751 \pm 0.014$ and $\rho_{11} = 0.252 \pm 0.014$, indicating that unintended population leakage is small within experimental uncertainty. Furthermore, the off-diagonal coherences are suppressed to $0.009 \pm 0.015$, with imaginary components consistent with zero within the shot-noise limit of 4500 repetitions [Fig.~\ref{fig:mixed_state}(b)]. These data support the view that state-preparation errors are not the dominant source of deviation in the thermodynamic measurements.

\section{Dynamics of the No-Jump Survival Probability}
\label{app:survival}

In this section, we detail the dynamics of the no-jump survival probability $S_i(T)$, which sets the experimental postselection efficiency of our passive non-Hermitian framework. As governed by the effective non-Hermitian Hamiltonian, the macroscopic population undergoes a continuous decay. This loss is primarily driven by the global anti-Hermitian dissipation term $-i\gamma\mathbb{I}$, which imposes an overall exponential decay envelope proportional to $e^{-2\gamma t}$. Superimposed on this baseline are coherent oscillations arising from the non-unitary nature of the dynamics.

\begin{figure}[H]
    \centering
    \includegraphics[width=8cm]{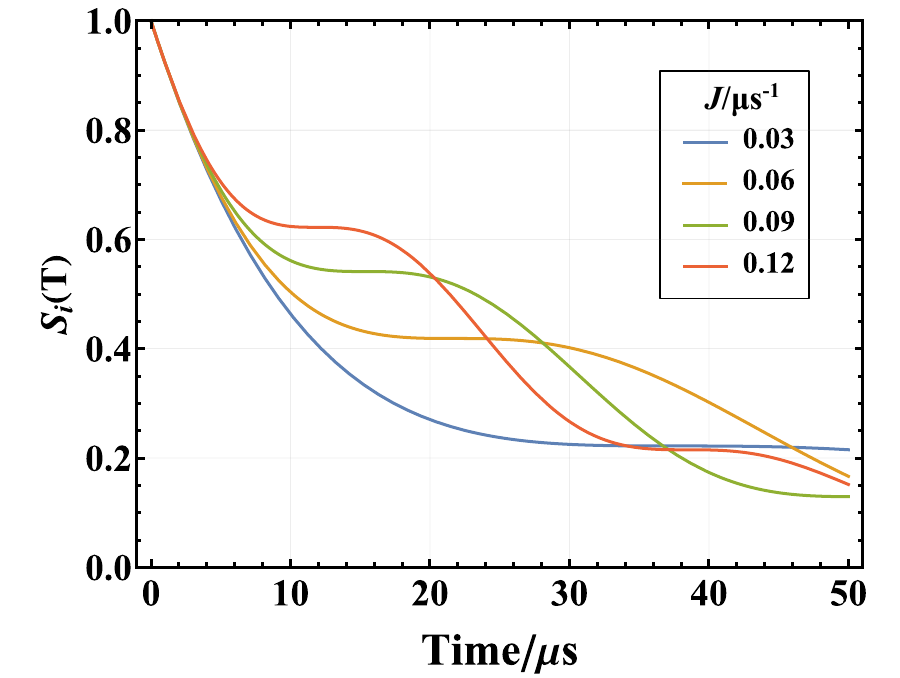}
    \caption{\textbf{Continuous-evolution theoretical postselection cost, quantified by the dynamical decay of the no-jump survival probability $S_i(T)$.} Solid lines represent the simulated continuous survival probabilities under the effective non-Hermitian dynamics for various driving strengths $J$, ranging from $0.03\,\mu\text{s}^{-1}$ to $0.12\,\mu\text{s}^{-1}$. While higher driving strengths induce more rapid non-unitary oscillations, the baseline exponential decay governed by the macroscopic dissipation rate $\gamma=0.02\,\mu\text{s}^{-1}$ results in a rapid decay of the survival probability across all conditions.}
    \label{fig:survival}
\end{figure}

As illustrated in Fig.~\ref{fig:survival}, the curves represent the continuous-evolution theoretical postselection cost. While the specific frequency of these oscillations varies with the driving strength $J$, the survival probability $S_i(T)$ shows a rapid decay across all driving conditions. This population decay highlights the substantial postselection cost inherent to passive non-Hermitian platforms. Because the main measurements use a piecewise reinitialization protocol, the measured count rate is not 
identical to the continuous survival probability in Fig.~\ref{fig:survival}. We therefore report Fig.~\ref{fig:survival} as the theoretical continuous-evolution postselection cost, while the piecewise protocol is used only to reconstruct the 
conditional transition dynamics with improved signal-to-noise ratio.

\section{Extraction of the Effective Floquet Hamiltonian}
\label{app:floquet}
In this section, we detail the algebraic and numerical procedure used to extract the effective Floquet Hamiltonian $H_{\text{F}}$ from the non-unitary evolution operator. This methodology explicitly avoids the branch-cut ambiguities inherent to multi-valued matrix logarithms and identifies the criteria for the revival of the Jarzynski equality under an arbitrary hybridization angle $\theta_k$.

We begin by isolating the deterministic macroscopic decay from the coherent dynamics. Consider the total time-evolution operator over one driving period $T$, denoted as $U'(\theta_k,T)$ [see Eq.~\eqref{eq:U'}]. Because the global dissipation term $-i\gamma \mathbb{I}$ commutes with any generic SU(2) rotation, the total operator can be factored as:
\begin{equation}
    U'(\theta_k,T) = U_r(\theta_k,T) e^{-\gamma T},
\end{equation}
where $U_r(\theta_k,T)$ is the relative non-unitary evolution operator. This factorization ensures that the macroscopic survival probability does not obscure the symmetry structure underlying the non-Hermitian transitions. Because the relative operator preserves a unit determinant ($\det[U_r(\theta_k,T)] = 1$), the corresponding relative Floquet Hamiltonian $H_{\text{F}}(\theta_k,T)$ is traceless and can be expressed via Pauli matrices as $H_{\text{F}}(\theta_k,T) = \vec{h}(\theta_k,T) \cdot \vec{\sigma}$. Here, $E_{\text{F}}(\theta_k,T) = \sqrt{\vec{h}(\theta_k,T) \cdot \vec{h}(\theta_k,T)}$ defines the complex quasi-energy of the system. 

To evaluate $H_{\text{F}}(\theta_k,T)$ while avoiding the ambiguities of multi-valued numerical matrix logarithms, we define the dynamic phase $\theta(T) = E_{\text{F}}(\theta_k,T) T$ and analytically expand the evolution operator using $SU(2)$ algebra: $U_r(\theta_k,T) = \cos\theta(T)\mathbb{I} - i \frac{\sin\theta(T)}{\theta(T)} \big(H_{\text{F}}(\theta_k,T) T\big)$. This complex phase is explicitly constrained by the trace of the evolution operator:
\begin{equation}
    \theta(T) = \arccos\left( \frac{1}{2}\text{Tr}[U_r(\theta_k,T)] \right).
\end{equation}
To resolve the infinite branch choices associated with the arccosine function, we enforce dynamical continuity. By continuously unwrapping the phase $\theta(T)$ from the static limit $T \to 0$ (where $\theta(0) = 0$) along finely discretized time steps, we extract the physically continuous phase. Substituting this unique $\theta(T)$ back yields the reconstruction of the relative Floquet Hamiltonian:
\begin{equation}
    H_{\text{F}}(\theta_k,T) = \frac{i\theta(T)}{T\sin\theta(T)}\Big(U_r(\theta_k,T) - \cos\theta(T)\mathbb{I}\Big).
\end{equation}

To map the dynamics back to the generalized initial state parameterized by $\theta_k$, we project the extracted relative Floquet Hamiltonian $H_{\text{F}}(\theta_k,T)$ onto the rotated hybrid basis. The Hermitian driving axis $\sigma_{\parallel}$ and the orthogonal dissipative axis $\sigma_{\perp}$ are defined as:
\begin{align}
    \sigma_{\parallel}(\theta_k) &= \sin\theta_k \sigma_x - \cos\theta_k \sigma_z, \\
    \sigma_{\perp}(\theta_k) &= \cos\theta_k \sigma_x + \sin\theta_k \sigma_z.
\end{align}
The effective Floquet parameters are thus identified by their corresponding projections: the generalized driving strength $J_{\text{eff}}(\theta_k,T) = \text{Re}[h_{\parallel}(\theta_k,T)]$, the relative dissipation $\gamma_{\text{eff}}(\theta_k,T) = \text{Im}[h_{\perp}(\theta_k,T)]$, and the orthogonal Hermitian deflection $\text{Re}[h_y(\theta_k,T)]$.

For the Jarzynski equality to revive, the Hermitian part of $H_{\text{F}}(\theta_k,T)$ must realign with the initial generalized basis \cite{Erdamar2024}, while the dissipative structure remains orthogonal to the driving axis $\sigma_{\parallel}$. Algebraically, this  revival requires four specific Floquet components to vanish simultaneously: the in-plane and out-of-plane Hermitian deflections ($\text{Re}[h_{\perp}(\theta_k,T)]$ and $\text{Re}[h_y(\theta_k,T)]$), alongside the non-orthogonal dissipations ($\text{Im}[h_{\parallel}(\theta_k,T)]$ and $\text{Im}[h_y(\theta_k,T)]$). By tracking their dynamical trajectories, we identify the symmetry revival times $\tau_{\text{rev}}$ via the joint root-finding condition:
\begin{equation}
    \begin{cases} 
    \text{Re}[h_{\perp}(\theta_k, \tau_{\text{rev}})] = 0, & \text{Re}[h_y(\theta_k, \tau_{\text{rev}})] = 0 \\
    \text{Im}[h_{\parallel}(\theta_k, \tau_{\text{rev}})] = 0, & \text{Im}[h_y(\theta_k, \tau_{\text{rev}})] = 0
    \end{cases}
\end{equation}
At these revival times, the Floquet Hamiltonian reduces to the canonical hybrid form $H_{\text{F}}(\theta_k, \tau_{\text{rev}}) = J_{\text{eff}}(\theta_k,\tau_{\text{rev}})\sigma_{\parallel}(\theta_k) + i\gamma_{\text{eff}}(\theta_k,\tau_{\text{rev}})\sigma_{\perp}(\theta_k)$, ensuring that the transition probabilities obey the symmetry-enforced thermodynamic relation.
\section{High-Resolution Temporal Scans Near Symmetry Revival Points}
\label{app:finescan}

To further examine the symmetry revival points $T_1$ and $T_2$ and provide a consistency check on our piecewise protocol, we performed high-resolution temporal fine scans. We choose the $\mathcal{PT}$-symmetric case as a representative example to resolve the local behavior near the predicted crossings.

\begin{figure}[H]
    \centering
    \includegraphics[width=8cm]{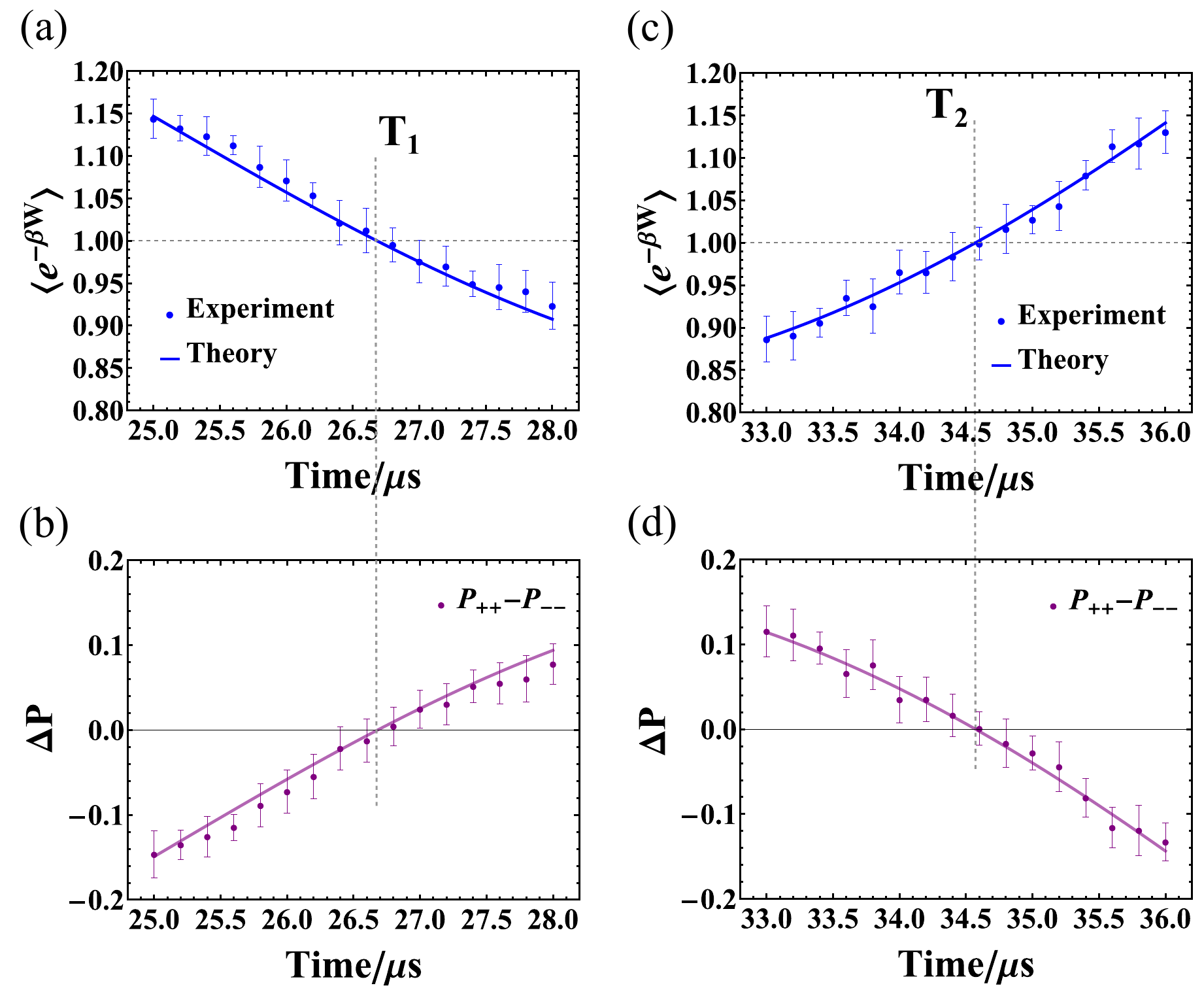} 
    \caption{High-resolution fine scans near the symmetry revival points for the $\mathcal{PT}$-symmetric configuration. Experimental data (symbols with error bars) are sampled with a fine step size of $0.2\,\mu\text{s}$. (a) and (c) Evolution of the exponential work $\langle e^{-\beta W} \rangle$ near $T_1$ and $T_2$, respectively. (b) and (d) Corresponding dynamic tracking of the transition probability difference $\Delta P = P_{++} - P_{--}$. The vertical dashed lines indicate the theoretical revival times. At these points, $\Delta P$ exhibits a zero crossing, signaling the restoration of parity-exchange symmetry, which in turn drives the macroscopic exponential work to cross the $\langle e^{-\beta W} \rangle = 1$ baseline.}
    \label{fig:finescan_appendix}
\end{figure}

As shown in Fig.~\ref{fig:finescan_appendix}, we tracked the evolution within two high-resolution windows: $T \in [25, 28]\,\mu\text{s}$ and $T \in [33, 36]\,\mu\text{s}$. With a temporal resolution of $\Delta t = 0.2\,\mu\text{s}$, the experimental data trace the theoretical zero-crossing contours. The measured $\Delta P$ crosses zero near  $T_1 \approx 26.7\,\mu\text{s}$ and $T_2 \approx 34.6\,\mu\text{s}$. These zero-crossings, observed with small statistical fluctuations, provide an additional consistency check on the initialization fidelity and stability of our digital evolution scheme. This supports the view that the conditional symmetry-based thermodynamic relation holds within experimental resolution in our non-Hermitian platform.
\section{Experimental Results for the Intermediate Angle $\theta_k = \pi/4$}
\label{app:pi4}
In this section, we present additional experimental results for the intermediate hybridization angle $\theta_k = \pi/4$ to test 
the theoretical symmetry pattern away from the two endpoint models.
\begin{figure}[H]
    \centering
    \includegraphics[width=8cm]{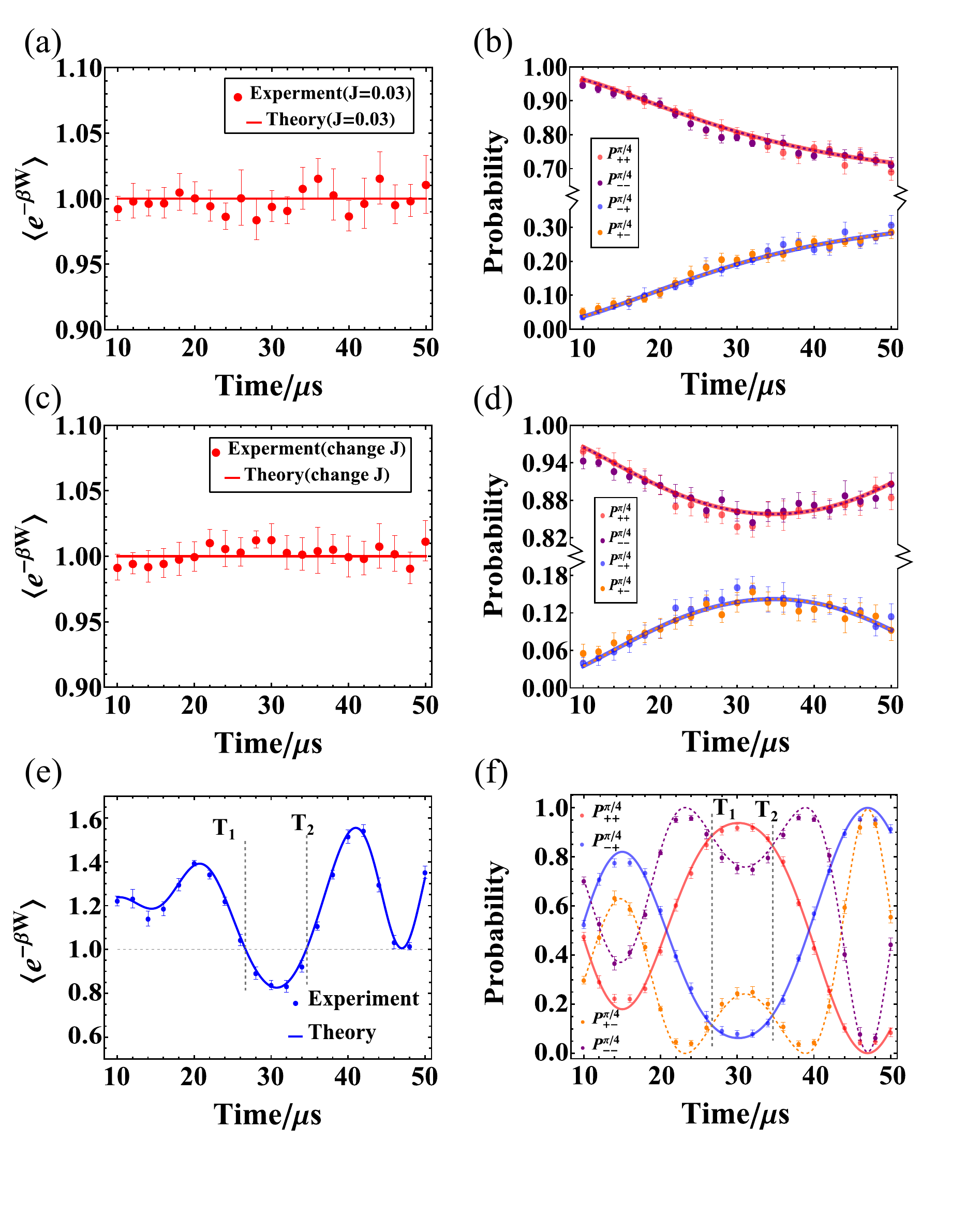}
    \caption{Experimental and theoretical results for the maximally hybridized intermediate angle $\theta_k = \pi/4$. The data points with error bars represent experimental results, while the solid or dashed lines denote theoretical results. (a) and (b) Exponential work and transition probabilities under a constant coupling $J = 0.03\,\mu\text{s}^{-1}$. (c) and (d) Results under a linearly varying coupling $J(t)$. (e) and (f) Results with a sinusoidally modulated detuning added to the Hamiltonian. The evolution times $T_1$ and $T_2$ denote the instances where the effective Floquet Hamiltonian regains the targeted symmetry and the exponential work returns to 1.}
    \label{fig:pi4}
\end{figure}
Fig.~\ref{fig:pi4} displays the exponential work $\langle e^{-\beta W} \rangle$ and transition probabilities under the three temporal driving protocols discussed in the main text. Specifically, Figs.~\ref{fig:pi4}(a) and \ref{fig:pi4}(b) correspond to the evolution with a constant coupling $J = 0.03\,\mu\text{s}^{-1}$ [Eq.~\eqref{eq:const}]. The results show that the relation $\langle e^{-\beta W} \rangle \approx 1$ is satisfied within experimental uncertainty. As predicted by the operator-level geometric analysis, the measured transition probabilities exhibit the parity-exchange symmetry ($P_{++}^{\pi/4} = P_{--}^{\pi/4}$ and $P_{+-}^{\pi/4} = P_{-+}^{\pi/4}$) expected by the theory.

Fig.~\ref{fig:pi4}(c) and \ref{fig:pi4}(d) present the case where $J(t)$ varies linearly from $0.03\,\mu\text{s}^{-1}$ to $0.06\,\mu\text{s}^{-1}$ and then returns to $0.03\,\mu\text{s}^{-1}$ [Eq.~\eqref{eq:change}]. Similar to the static case, the transition probabilities remain consistent with parity-exchange symmetry, supporting the validity of the Jarzynski equality for this protocol.

Finally, in Fig.~\ref{fig:pi4}(e) and \ref{fig:pi4}(f), we introduce the sinusoidally modulated detuning [Eq.~\eqref{eq:dchange}] into the hybrid Hamiltonian. Consistent with the pure $\mathcal{PT}$ and $\mathcal{APT}$ scenarios, the time-dependent detuning generally breaks the parity-exchange symmetry for most of the evolution time, leading to a violation of the Jarzynski equality. However, at the specific evolution times $T_1$ and $T_2$ where the effective Floquet Hamiltonian recovers the specific hybrid Hamiltonian form, the parity-exchange symmetry of the transition probabilities is regained. At these instances, the exponential work returns 
to a value consistent with 1 within experimental uncertainty.

Collectively, these intermediate-angle results support the robustness of our experimental implementation and are consistent with the theoretical claim that the parity-exchange criterion is preserved along the $\mathrm{SU}(2)$-rotated family in this two-level setup.

\section{Statistical Analysis and Error Estimation}
For the experimental data, conditional transition probabilities are extracted via a piecewise evolution scheme with postselection, normalizing state readouts to the surviving population within the qubit subspace. Each data point comprises 4,500 repetitions divided into 15 independent temporal blocks, yielding measured standard deviations less than 0.025. The state preparation fidelity is above 99\%, indicating that state preparation and measurement (SPAM) errors are maintained at a low baseline level. These remnant fluctuations are not primarily limited by statistical sampling or SPAM error; instead, they reflect the physical variations present in the trapped-ion setup, where residual low-frequency drifts perturb the effective Hamiltonian parameters. Specifically, slow power drifts in the microwave driving field affect the coherent coupling strength ($J$), and intensity instabilities of the optical dissipation beam introduce fluctuations in the non-Hermitian decay rate ($\gamma$). Consequently, the reported error bars represent the physical uncertainties associated with these system drifts.

\bibliographystyle{unsrt} 

\bibliography{references}
\end{document}